\documentclass[aps,prl,floatfix,twocolumn]{revtex4}

\usepackage{amsmath}
\usepackage{epsfig}
\usepackage{hyperref}

\begin{document}

\title{Individual addressing of ions using magnetic field gradients \\ in a surface-electrode ion trap}

\author{Shannon X. Wang}
\email[]{sxwang@mit.edu}
\author{Jaroslaw Labaziewicz}
\author{Yufei Ge}
\author{Ruth Shewmon}
\author{Isaac L. Chuang}
\affiliation{Massachusetts Institute of Technology, Center for Ultracold Atoms, Department of Physics, 77 Massachusetts Avenue, Cambridge, MA, 02139, USA}

\date{\today}

\begin{abstract}
\noindent
Dense array of ions in microfabricated traps represent one possible way to scale up ion trap quantum computing. The ability to address individual ions is an important component of such a scheme. We demonstrate individual addressing of trapped ions in a microfabricated surface-electrode trap using a magnetic field gradient generated on-chip. A frequency splitting of 310(2)~kHz for two ions separated by 5~$\mu$m is achieved. Selective single qubit operations are performed on one of two trapped ions with an average of 2.2$\pm$1.0\% crosstalk. Coherence time as measured by the spin-echo technique is unaffected by the field gradient.
\end{abstract}

\maketitle


Trapped ion systems are promising candidates for implementing quantum computation \cite{Wineland:98}. Significant progress has been made in realizing quantum operations using a small number of ions as qubits \cite{SchmidtKaler:03,Riebe:04,Barrett:04,Chiaverini:04a,Blatt:08}. Meaningful quantum computation involves a large number of qubits, and one possible way to scale up ion traps is to create a dense array of ions in microfabricated traps. The ability to individually address ions is an essential component of such a system. In this letter, we report the successful demonstration of individual addressing of single trapped ions in a linear crystal, using a magnetic field gradient generated in a microfabricated surface-electrode trap.

Previous methods of individual ion addressing have included spatial and frequency separation. An example of spatial separation is the use of precisely focused laser beams aimed at only one ion at a time \cite{Nagerl:99}, which poses a significant technical challenge in laser beam control. Another approach is to transport ions between trap zones using varying DC potentials \cite{Barrett:04,Leibfried:07}, which requires precision voltage and timing control. Separation in frequency space has been proposed and implemented by creating AC stark shifts using a far off-resonant laser \cite{Staanum:02,Haljan:05}. Frequency selectivity can also be achieved by applying an inhomogeneous magnetic field gradient for spatially-separated ions \cite{Wunderlich:01}. This has been demonstrated in a linear Paul trap using hyperfine states of trapped $^{172}$Yb$^+$ ions probed with rf radiation \cite{Wunderlich:08}, and with neutral Cs atoms in an optical dipole trap \cite{Schrader:04}. A similar technique can be applied to magnetic-field-sensitive qubit transitions in the optical range. Such an approach requires separating the ion frequencies by much more than the desired gate speed \cite{Wineland:98}. 

Here, we present the first design and implementation of a microfabricated trap to perform individual addressing for optical transitions using a magnetic field gradient. Addressability is demonstrated by observing distinct peaks in the frequency spectrum, and by performing Rabi oscillations on one of two trapped ions with a low probability of unwanted excitation on the neighbouring ion. Integrating the gradient-generating wires with microfabricated traps assures position stability. We evaluate this individual addressing scheme to show that we can achieve reasonable gate speeds, and minimal crosstalk between neighbouring ions, with no decrease in coherence time measured with spin-echo. Finally we consider its prospects for scaling to many ions.

\begin{figure}[b]\includegraphics{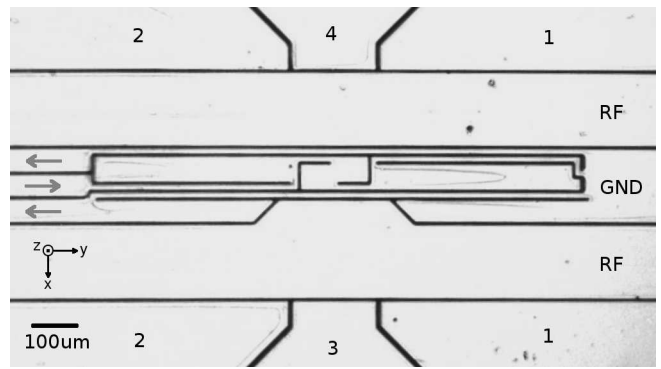}
\caption{\label{Fig:gradient} Microscope image of the microfabricated trap. Light regions are electroplated gold, and dark regions are exposed dielectrics (quartz). Gray arrows indicate current paths. DC electrodes are numbered 1-4. }
\end{figure}

The gradient trap is based on a five-rod surface-electrode design similar to that of Ref. \cite{Labaziewicz:08}, with the trap center 100~$\mu$m above the surface. The fabrication process is identical to that of Ref.\cite{Labaziewicz:08b}. A single-layer design simplifies microfabrication, as well as enables proximity between the gradient-generating current and the ion. The ground electrode is patterned as shown in Fig. \ref{Fig:gradient}. Three parallel current paths are created. The center current path follows an S-shaped pattern directly below the trap center, and splits into two paths at the opposite end of the electrode. This symmetry is necessary to minimize the horizontal bias field in the axial direction. The narrowest path width is 10~$\mu$m. The dimensions of the current path geometry are chosen to maximize the field gradient and minimize residual field at the trap center. Calculations using the Biot-Savart law were performed to optimize this geometry. The trap is operated in a bath cryostat cooled to 4~K, at which temperature the total resistance of conducting path on the trap is 0.2~$\Omega$. Up to 500~mA of current is applied, dissipating 50~mW of power on the trap. Ions are loaded by photoionization of a thermal vapor, and the initial loading rate is slow enough to allow deterministic loading of 1-3 ions \cite{Balzer:06}. 

The \mbox{$5$S$_{1/2}(m=-1/2)\rightarrow4$D$_{5/2}(m=-5/2)$} optical transition of $^{88}$Sr$^+$ ions at 674~nm forms the qubit, chosen for the long 345~ms lifetime \cite{James:98a} of the metastable 4D$_{5/2}$ state. The degeneracy of multiple Zeeman transitions is lifted by applying a constant bias field of about 4~G using external coils. The separation of the qubit frequencies of two ions spaced by s in a field gradient $\partial B_z/\partial y$ is given by $f = \delta g \mu_B s \partial B_z/\partial y$ ($z$ is the quantization axis and $y$ is along the direction of the ion chain), where $\mu_B$ is the Bohr magneton and $\delta g=2$ is the difference in Land\'e g-factors for the S and D levels. This separation needs to be much larger than the Rabi frequency to reduce crosstalk. For ion spacing $\sim$5~$\mu$m and Rabi frequency $\sim$100~kHz, a gradient of $\gg$~7.2~G/mm is needed.

\begin{figure}
\includegraphics{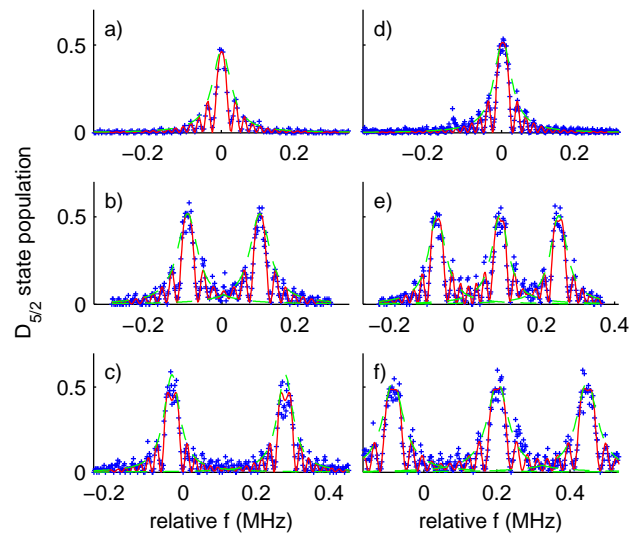} 
\caption{\label{Fig:spectra} (color online) Spectroscopy on the $5$S$_{1/2}\rightarrow4$D$_{5/2}$ transition of $^{88}$Sr$^+$ for (a-c) two and (d-f) three ions, with (a,d) 0~mA, (b,e) 300~mA, and (c,f) 500~mA of current applied to the gradient structure. Each spectrum is fitted to the sum of Lorentzians with $\sin^2$ modulation (solid red lines), the Rabi resonance transition probability. Frequencies are relative to the peak centers in (a,d). Dashed green lines are envelopes of the individual Lorentzians at each frequency. Averaged frequency splittings with 500~mA are 310(2)~kHz for two ions and 266(1)~kHz for three ions. }
\end{figure}

The spectrum of the qubit transition, shown in Fig. \ref{Fig:spectra}, demonstrates individual addressing in frequency space. For these measurements, the ions are first Doppler cooled to the 5S$_{1/2}$ state, then the frequency of the 674~nm laser is scanned across the qubit transition. At each frequency, a 50~us pulse is applied, and transition events are detected by applying a readout pulse tuned to the S-P transition and collecting the fluorescence. This measurement is repeated 100 times for each frequency to determine the transition probability. The resulting spectrum for two and three ions at 0, 300~mA and 500~mA of applied current can be clearly resolved. We define crosstalk as the probability that ion 2 makes a transition given that a $\pi$ pulse is applied to ion 1. For the case of two ions at 300 mA, fitting the spectra to a sum of modulated Lorentzians bounds crosstalk to be below 2.8\% at Rabi frequency of 34~kHz. 

The actual field gradient can be determined from the ion spacing and the frequency splitting. The secular frequency for this experiment was determined to be 847~kHz, by measuring the frequency difference between the carrier and first sideband of the qubit transition.  This gives ion spacing of 4.8~$\mu$m for two ions and 4.1~$\mu$m for three ions. Together with the frequency splittings, this indicates a gradient of 14 G/mm at 300 mA and 23 G/mm at 500 mA. At the maximum applied current of 500 mA, the average frequency splittings are 310(2)~kHz for two ions and 266(1)~kHz for three ions. The uneven splitting for three ions is likely due to an inhomogeneous field gradient, which may be caused by asymmetric current paths. The gradient current also creates a small bias field (on the order of 10mG) which shifts the center frequency. 

\begin{figure}
\includegraphics{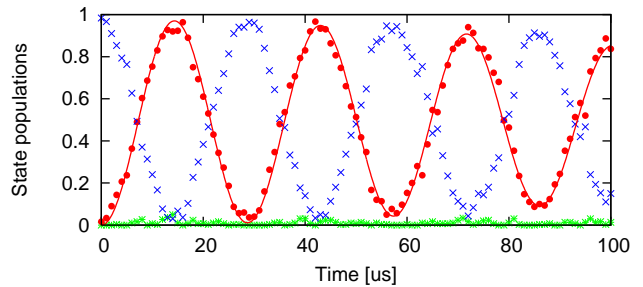}
\caption{\label{Fig:2ionRabi} 
Rabi flops on one of the two trapped ions separated in frequency by 190~kHz. Plotted are the D$_{5/2}$ state populations of two ions for the case of of no transitions ($\times$), of single-ion transition events (\textbullet), and of simultaneous transition events($*$). The fitted Rabi frequency to the data of single-ion transition events is 35~kHz, with a Gaussian envelope of HWHM of 170~$\mu$s and initial contrast of 97\%. }
\end{figure}

Rabi oscillations are performed on one of the two trapped ions at 190~kHz frequency splitting. Fig. \ref{Fig:2ionRabi} shows several Rabi oscillations using the higher of the two split frequencies. State population is measured by collecting the total fluorescence using a PMT, which does not distinguish the two ions.  Discriminating between two and one bright ion requires 1 ms of total integration time. The frequency of these oscillations is 35kHz, corresponding to 14~$\mu$s for a $\pi$-gate. A measurement of crosstalk can be obtained by applying a $\pi$ pulse and measuring the number of simultaneous transitions. The Rabi data gives a crosstalk of 2.2$\pm$1.0\%, within error of the expected value. 


To investigate possible effects on coherence time caused by the field gradient, we perform Ramsey spectroscopy on a single trapped ion with and without the gradient current. An upper bound on the coherence time $T^*_2$ is obtained by measuring the decay of Ramsey fringes as a function of the separation of the Ramsey $\pi/2$ rotations \cite{Sinclair:04}. In the absence of a field gradient, the envelope of the observed Ramsey oscillations follows a Gaussian decay with $T^*_2 = 632 \pm 12 \mu$s. With 300~mA of current applied to the gradient structure, $T^*_2 = 424 \pm 9 \mu$s. However, by using spin-echo methods and applying a $\pi$ pulse at 0.5~ms and subsequently every 1~ms, an exponentially decaying fit to the peak of recovered Ramsey oscillations has a time constant of $T_2\sim$10~ms for both cases. There are several possible explanations for the source of decoherence, such as noise in the current source, increased sensitivity to horizontal magnetic field fluctuations due to a horizontal bias field, and ion movement. Some of these can be removed, e.g. by inductively filtering the current input, or applying a horizontal field to cancel the horizontal bias. 
While further work is needed to identify and remove the inhomogeneous sources of noise, it's removed by spin-echo and the decoherence effects of the gradient become negligible. The spin-echo result indicates that the noise added by the gradient is low-frequency (less than 1~kHz). 


Scaling up this scheme to N~$\ge$~3 ions would require a larger field gradient, since ion spacing decreases as N increases \cite{James:98a}. For cryogenic systems, the gradient current is limited by power dissipation. A superconducting version of the trap with a persistent current loop would enable higher currents and gradient with negligible power dissipation. Also, multi-layer fabrication would place fewer constraints on the design of the current path, allowing the same current to generate a much larger field gradient. An issue which needs addressing is the phase shifts created by the near resonant light fields \cite{Wineland:98}. Since frequency splittings are known, these phase shifts can be corrected by controlling the phase of the addressing laser.


In conclusion, we have demonstrated individual addressing of trapped ions using a magnetic field gradient, generated with microfabricated structures in a surface-electrode trap. Separation in frequency space is demonstrated for up to three ions, and Rabi flops on one of two ions are performed with an average of 2.2$\pm$1.0\% crosstalk. Furthermore, the coherence time of a single ion is unaffected by the magnetic field gradient when spin echo pulses are used to remove low-frequency noise. These features make this scheme an attractive alternative for implementing individual addressing of ions for many-qubit quantum information processing. Future work will include the fabrication of the gradient structure in a superconducting trap as a persistent current loop to create greater gradient in a cryogenic setup. Meanwhile, implementing addressing error correction with composite pulses \cite{Aolita:07} would enable higher gate speeds. 

This work was supported by the Japan Science and Technology Agency and the NSF Center for Ultracold Atoms.


\end{document}